\documentclass[twocolumn,showpacs,preprintnumbers,showkeys,superscriptaddress]{revtex4}
\usepackage{CJK}
\usepackage{multirow}
\usepackage{mathrsfs}
\usepackage{amssymb}
\usepackage{amsfonts}
\usepackage{amsmath}
\usepackage{graphicx}
\usepackage{dcolumn}
\usepackage{bm}
\usepackage{float}
\usepackage[normalem]{ulem}
\usepackage{color}

\begin{document}
\begin{CJK*}{UTF8}{} 
\title{Robust upper limit on the neutron single-particle energy of the $i_{13/2}$ orbit}

\author{Y. Lei ({\CJKfamily{gbsn}雷杨})}
\email{leiyang19850228@gmail.com}
\affiliation{Key laboratory of neutron physics, Institute of Nuclear Physics and Chemistry, China Academy of Engineering Physics, Mianyang 621900, China}

\author{H. Jiang ({\CJKfamily{gbsn}姜慧})}
\affiliation{School of Arts and Sciences, Shanghai Maritime University, Shanghai 201306, China}
\affiliation{Department of Physics, Shanghai Jiao Tong University, Shanghai 200240, China }

\date{\today}
\begin{abstract}
The upper limit of the neutron $i_{13/2}$ single-particle energy ($\varepsilon_{i13/2}$) is estimated in the random quasiparticle ensemble with single-particle energies. The $\varepsilon_{i13/2}$ distributions under constraints from $^{134,~135}$Te	and	$^{136,~137}$Xe spectra demonstrate that the $i_{13/2}$ single-particle state can be physically mixed with the $f_{7/2}\otimes 3^−$ configuration only if $\varepsilon_{i13/2}<3$ MeV. Thus, our ensemble calculation suggests a robust upper limit of 3 MeV.
\end{abstract}
\maketitle
\end{CJK*}

Neutron $i_{13/2}$ single-particle (s.p.) energy is a basic input of microscopic-model calculation for neutron-rich nuclei beyond the $^{132}$Sn core ($Z = 50,~N = 82$). It is also essential to calculate the life time of the $\beta$-decay $\nu i_{13/2}\rightarrow \pi h_{11/2}$ process, and the neutron emission probability of Sn resonant states, which both affect the final abundance of $A\sim 132$ elements from the $r$-process \cite{1}. However, the $i_{13/2}$ s.p. level is still unobserved in the experimental $^{133}$Sn level scheme, and supposed to be above the neutron-separation energy of $^{133}$Sn. Therefore, many efforts have been devoted to evaluate the $i_{13/2}$ s.p. energy (denoted by $\varepsilon_{i13/2}$) theoretically or experimentally. For instance, Refs. \cite{2,3} predicted the $i_{13/2}$ level above 3.5 MeV with the Nilsson model and relativistic mean-field theory. Based on experimental spectra of $^{134}$Sb and $^{210}$Bi, Ref. \cite{4} estimated $\varepsilon_{i13/2}=2.694(200)$ MeV, consistent with the prediction from Ref. \cite{5}. To provide valuable constraints on	$\varepsilon_{i13/2}$, $3^-_1$ and $13/2^+_1$ excitation	 energies (denoted by $E_{3-}$ and $E_{13/2+}$) of $^{134,~135}$Te and $^{137}$Xe were measured recently \cite{6}.

The purpose of this work is to further probe $\varepsilon_{i13/2}$ under constraints proposed by Ref. \cite{6} by using exact shell-model calculations \cite{7}. Conventionally, one can optimize $\varepsilon_{i13/2}$ by fitting the shell-model output to experimentally observed levels. However, this fitting process depends on the adopted two-body effective interaction, which introduces potential bias, and thus may deviate from physical reality. Therefore, we do not seek to determine $\varepsilon_{i13/2}$ explicitly for some specific effective interaction, but to estimate a robust limit of	$\varepsilon_{i13/2}$, which is compatible with most possibilities of reasonable two-body interactions. Random interaction is introduced to represent these interaction possibilities in this work. A single run of shell-model calculation with random interaction produces one ``sample" in the random-interaction ensemble. In a statistic point of view, a large number of samples under experimental constraints could provide a robust $\varepsilon_{i13/2}$ limit with physical consideration. It is noteworthy that the random-interaction ensemble was mainly used to investigate robust properties of generic many-body systems, e.g., the spin-zero groundstate dominance \cite{8,9} and the predominance of collective motions \cite{10,11,12,13}; see Refs. \cite{14,15,16,17} for reviews. It is a new attempt to study a realistic and specific nuclear problem (e.g., the limit of $\varepsilon_{i13/2}$ here) with random interaction. 

\begin{table}
\caption{Single-particle energies adopted in this work. Experimentally available s.p. energies are from Refs. \cite{18,19}. $\varepsilon_{i13/2}$ is undetermined (marked by ``$-$").}\label{tablei}
\begin{tabular}{cccccccccccccccccccccccccccccccc}
\hline\hline
\multirow{2}{*}{$\varepsilon_{\pi}$ (MeV)} 	&	$s_{1/2}$	&	$d_{3/2}$	&	$d_{5/2}$	&	$g_{7/2}$	&	$h_{11/2}$	&		\\
	&	2.990	&	2.440	&	0.962	&	0.000	&	2.792	&		\\
\hline
\multirow{2}{*}{$\varepsilon_{\nu}$ (MeV)}&	$p_{1/2}$	&	$p_{3/2}$	&	$f_{5/2}$	&	$f_{7/2}$	&	$h_{9/2}$	&	$i_{13/2}$	\\
	&	1.363	&	0.853	&	2.004	&	0.000	&	1.561	&	$-$	\\
\hline\hline
\end{tabular}
\end{table}

\begin{figure}
\includegraphics[angle=0,width=0.48\textwidth]{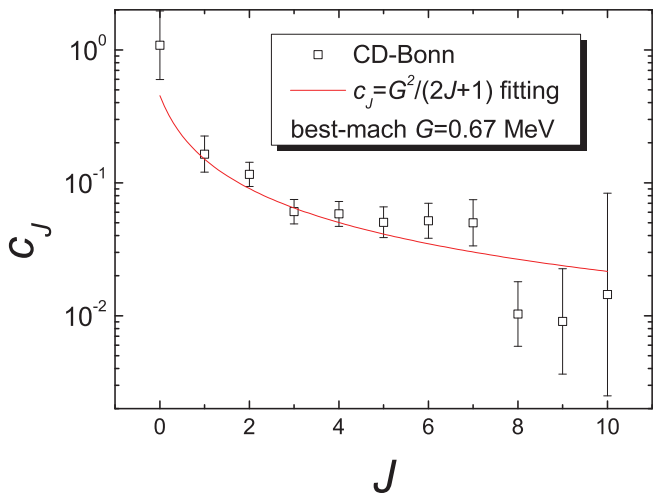}
\caption{(Color online) $c_J-J$ relation for realistic CD-Bonn two-body matrix elements. The black squares represent the calculated results from the width statistics of CD-Bonn elements, and corresponding error is estimated by the numerical experiment of Gaussian random number sampling. The red solid lines represent the best-match $c_J=G^2/(2J + 1)$ relation with $G=0.67$ MeV.}\label{fig1}
\end{figure}

In our calculations, the general shell-model Hamiltonian without isospin conservation is introduced, including s.p. energies from experiments and random two-body interaction. Adopted s.p. energies are listed in Table \ref{tablei}. The random two-body matrix element is denoted by $V_{\alpha\beta}$, where $\alpha$ or $\beta$ represents an arbitrary two-body configuration as $J=j_1\otimes j_2$ with total spin $J$ and nucleons at $j_1$ and $j_2$ orbits. $\alpha$ and $\beta$ have	the same $J$ value due to angular-momentum conservation. To construct an invariant random-interaction ensemble under arbitrary orthogonal transformation, $V_{\alpha\beta}$ is a Gaussian random number with mean zero and width as
\begin{equation}\label{eq1}
\langle V_{\alpha\beta}V_{\gamma\kappa} \rangle=C_J\frac{1+\delta_{\alpha\beta}}{2}\delta_{\alpha\gamma}{\beta\kappa},
\end{equation}
where $c_J=G^2/(2J + 1)$ following the definition of the random-quasiparticle ensemble with s.p. energies (RQE-SPE) \cite{8}, and $G$ is the overall energy scale of the RQE-SPE. The $c_J=G^2/(2J + 1)$ relation insures the invariance of the RQE-SPE under the transformation from the particle-particle representation to the particle-hole one. $G$ is determined by matching the widths, i.e., $c_J$ in Eq. (\ref{eq1}), of realistic two-body matrix elements. We calculate $c_J$ of the realistic CD-Bonn interaction \cite{20} for each $J$, and present it in Fig. \ref{fig1}. The CD-Bonn $c_J$ is a decreasing function of $J$ similarly to the $c_J=G^2/(2J+1)$ relation of the RQE-SPE. This similarity implies the applicability of the RQE-SPE for the study of a realistic nuclear system, which is the reason to adopt the RQE-SPE in this work. We fit $c_J = G2/(2J + 1)$ to the CD-Bonn $c_J-J$ relation with $G$ as the fitting parameter.The best-match $G = 0.67$ MeV is obtained, and the resultant RQE-SPE $c_J -J$ relation (red line in Fig. \ref{fig1}) reasonably describes the statistic property of the realistic CD-Bonn interaction.

With the best-match $G = 0.67$ MeV, the limit of $\varepsilon_{i13/2}$ can be revealed by its ensemble distributions under constraints from experimental spectra. Calculations of these $\varepsilon_{i13/2}$ distributions can be described as four steps. We first generate shell-model Hamiltonians of the RQE-SPE with $G=0.67$ MeV. $\varepsilon_{i13/2}$ in these Hamiltonians is also randomized, and follows the 0-8 MeV average distribution to represent all the $\varepsilon_{i13/2}$ possibilities from 0 MeV to the major shell energy ($\hbar \omega= 41/A^{1/3}\sim 8$ MeV). Secondly, we perform exact shell-model calculations for $^{134,~135}$Te or $^{136,~137}$Xe with random Hamiltonians generated above, and produce a large number of samples in the RQE-SPE. Thirdly, we introduce three types of physical constraints to filter out spectrally unphysical samples:
\begin{itemize}
\item
[(i)] The basic constraint requires samples to reproduce the spin-parity combination of the ground state, and the first excitation energy within 0.2 MeV error. This constraint guarantees the basic justifiability of this work. The 0.2 MeV error was previously estimated by Ref. \cite{4} to account for uncertainties from nucleon-nucleon residual interactions. Therefore, it is also taken as the acceptable spectral deviation from experiments in this work.
\item
[(ii)] The $E_{3-}$ constraint requires the sample to reproduce $E_{3-}$ of $^{134}$Te or $^{136}$Xe within 0.2 MeV error.
\item
[(iii)] The $E_{13/2+}$ constraint requires the sample to reproduce $E_{13/2+}$ of $^{135}$Te and $^{137}$Xe within 0.2 MeV error.	The $E_{3-}$ and $E_{13/2+}$ constraints are introduced according to Ref. \cite{6}.
\end{itemize}
In the fourth step, we calculate four types of $\varepsilon_{i13/2}$ distributions under physical constraints defined above, including
\begin{itemize}
\item
[(0)] the distribution under the basic constraint denoted by ``$P_0$",
\item
[(1)] the distribution under the basic constraint and the $E_{3-}$ constraint denoted by ``$P_1$",
\item
[(2)] the distribution under the basic constraint and the $E_{13/2+}$ constraint denoted by ``$P_2$",
\item
[(3)] the distribution under all the three constraints denoted by ``$P_3$".
\end{itemize}
We collect about 10000 samples for each	$\varepsilon_{i13/2}$ distribution calculation.

\begin{figure}
\includegraphics[angle=0,width=0.48\textwidth]{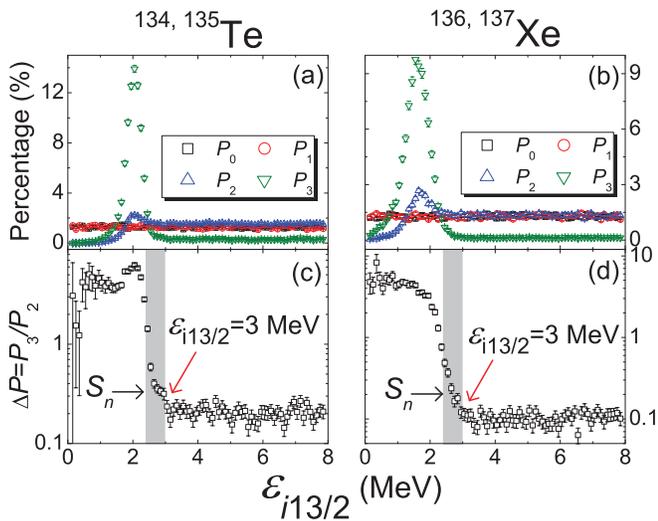}
\caption{(Color online) $\varepsilon_{i13/2}$ distributions. Panels (a) and (c) are for $^{134,~135}$Te calculations; panels (b) and (d) are for $^{136,~137}$Xe calculations. All the $\varepsilon_{i13/2}$ distributions in panels (a) and (b) are normalized, so that $\sum_{\varepsilon_{i13/2}}P(\varepsilon_{i13/2})=1$. The error in this figure is from sample counting. In panels (c) and (d), the increase of $\Delta P = P3/P2$ represents the mixing of the $i_{13/2}$ s.p. configuration and the $f_{7/2}\otimes 3^-$ configuration in the $13/2^+_1$ state of the $N = 83$ isotone (see text). The onset point of $\Delta P$ increase, i.e., the $\varepsilon_{i13/2}$ upper limit, is highlighted by red arrows and estimated as 3 MeV. Grey zones emphasize the $\varepsilon_{i13/2}$ range between the neutron separation energy of $^{133}$Sn ($S_n\sim 2.4$ MeV \cite{18,28,29,30}), and 3 MeV.}\label{fig2}
\end{figure}

Figures \ref{fig2}(a) and \ref{fig2}(b) present these four types of $\varepsilon_{i13/2}$ distributions	for $^{134,~135}$Te	and $^{136,~137}$Xe, respectively. $P_0$ is almost an average distribution, and thus can be a natural reference to investigate $\varepsilon_{i13/2}$. No background noise is introduced by the basic constraint. $P_1$ is also an average distribution under the pure $E_{3-}$ constraint. In other words, $\varepsilon_{i13/2}$ is not directly related to $E_{3-}$ of $N=82$ isotones. This behavior is consistent with the $N=82$ shell closure, which forbids the $i_{13/2}$ s.p. orbit to get involved in the $3^-$ excitation of $N = 82$ isotones.

Under the $E_{13/2+}$ constraint, both $P_2$ and $P_3$ present obvious peaks around $\varepsilon_{i13/2}\sim E_{13/2+}$ (2.109 and 1.725 MeV for Te and Xe, respectively), which corresponds to pure $i_{13/2}$ s.p. excitation with excitation energy equal to s.p. energy. It is the natural result of the RQE-SPE, which favors the pairing collectivity \cite{9,23}, and thus leaves the unpaired valence nucleon excited in the s.p. channel in odd-mass nuclei. On the other hand, $P_2$ and $P_3$ get trivialized around 0.5-1.5 \% in the $\varepsilon_{i13/2}=4-8$ MeV region, and follow the average distribution. This observation demonstrates that large $\varepsilon_{i13/2}$ will lose the impact on $E_{13/2+}$. We note that the $i_{13/2}$ systematic trend for $13/2^+_1$ states of $N = 83$ isotones are well established in the literature \cite{18}, and thus $E_{13/2+}$ shall be realistically related to $\varepsilon_{i13/2}$. Therefore, a $\varepsilon_{i13/2}$ upper limit should exist to enable the significant impact of $i_{13/2}$ s.p. excitation on $E_{13/2+}$.

We use the ratio of $P_3$ over $P_2$ (denoted by $\Delta P = P3/P2$) to estimate the $\varepsilon_{i13/2}$ upper limit as presented in Figs. \ref{fig2}(c) and \ref{fig2}(d). $\Delta P$ should be logically related to the $E_{3-}$ constraint, the only difference between $P_2$ and $P_3$ calculations. However, the $E_{3-}$ constraint has no direct effect on the $\varepsilon_{i13/2}$ distribution as suggested by $P_1$. Therefore, $\Delta P$ represents the coherence between $E_{3-}$ and $E_{13/2+}$ constraints. On the other hand, $E_{13/2+}$ of realistic $N = 83$ isotone is significantly affected by $\varepsilon_{i13/2}$ and $E_{3-}$, considering the physical mixing of the $i_{13/2}$ s.p.	 configuration and the $f_{7/2}\otimes 3^-$ configuration in the $13/2^+_1$ state \cite{24,25,26,27}. In other words, this mixing provides extra possibility to coherently fit both $E_{3-}$ and $E_{13/2+}$ constraints, and thus increases $\Delta P$ for physical	$\varepsilon_{i13/2}$. In the large-$\varepsilon_{i13/2}$ range, $\Delta P$ is a constant corresponding to trivially and averagely distributed $P_2$ and $P_3$, which demonstrates no physical mixing. In the small-$\varepsilon_{i13/2}$ range, $\Delta P$ significantly increases and represents the existence of the physical two-configuration mixing in the $13/2^+_1$ state. As shown in Figs. \ref{fig2}(c) and \ref{fig2}(d), the onset point of $\Delta P$ increase is identified as 3 MeV to trigger this mixing, and thus a physical $\varepsilon_{i13/2}$ upper limit is estimated to be 3 MeV.

\begin{figure}
\includegraphics[angle=0,width=0.48\textwidth]{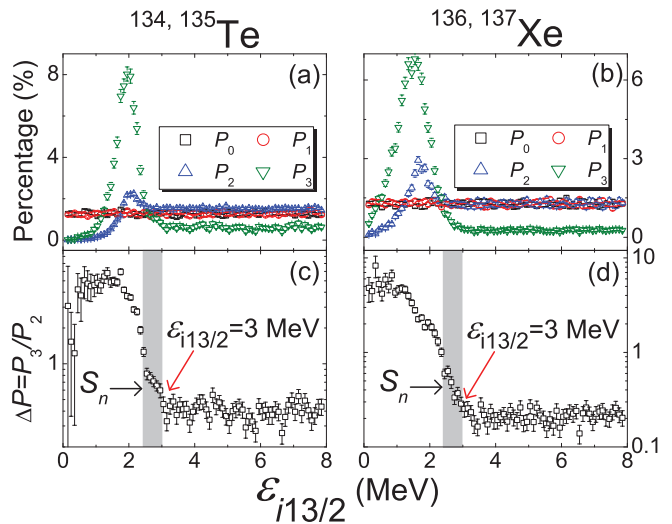}
\caption{(Color online) Similar to Fig. \ref{fig2} except with $G$ averagely distributed between 0 and 2 MeV. The $\varepsilon_{i13/2}$ upper limit is also estimated to be 3 MeV as in Fig. \ref{fig2}, which demonstrates the robustness of our results.}\label{fig3}
\end{figure}

To examine whether the $\varepsilon_{i13/2}$ upper limit is robust or sensitive to $G$, we also randomize $G$ averagely between 0 and 2 MeV, and repeat our calculation. Such calculated $P_0$, $P_1$, $P_2$, $P_3$, and $\Delta P$ are presented in Fig. \ref{fig3}. The $\varepsilon_{i13/2}$ distributions with random $G$ are similar to those with the best-match $G$ (see Fig. \ref{fig2}), and thus the $\varepsilon_{i13/2}$ upper limit with random $G$ is also estimated to be 3 MeV. Therefore, we conclude that the  $\varepsilon_{i13/2}$ upper limit from our RQE-SPE calculation is robust and insensitive to $G$ value.

Our RQE-SPE calculation seems to favor $\varepsilon_{i13/2}\sim E_{13/2+}$ peaks of $P_2$ and $P_3$ in Figs. \ref{fig2}(a), \ref{fig2}(b), \ref{fig3}(a), and \ref{fig3}(b). These peaks correspond to pure $i_{13/2}$ s.p. excitation as we comment above. However, the physical mixing of the $f_{7/2}\otimes 3^-$ configuration in the $13/2^+_1$ s.p. state will decrease $E_{13/2+}$, and thus leads to $E_{13/2+}<\varepsilon_{i13/2}$.	Therefore, $\varepsilon_{i13/2}\sim E_{13/2+}$	peak position is inconsistent with the physical configuration-mixing picture, and thus unreasonable. A strict $\varepsilon_{i13/2}$ lower limit is essential to exclude the unreasonable $\varepsilon_{i13/2}\sim E_{13/2+}$ peak region. We note that the $i_{13/2}$ s.p. level of $^{133}$Sn is potentially above the neutron separation energy (denoted by Sn), and thus the lower limit of $\varepsilon_{i13/2}>S_n\sim 2.4$ MeV \cite{18,28,29,30} can be introduced as an amendment of our SPE-RQE calculations. As a result, $\varepsilon_{i13/2}$ should lie in a narrow window between 2.4 and 3 MeV, as emphasized by grey zones in Figs. \ref{fig2}(c), \ref{fig2}(d), \ref{fig3}(c), and \ref{fig3}(d).

To summarize, the coherence of $E_{3-}$ and $E_{13/2+}$ constraints favors $\varepsilon_{i13/2}<3$ MeV range in our RQE-SPE calculations with both best-match and random $G$ values. This observation corresponds to the realistic mixing of the proton $3^-$ excitation and $i_{13/2}$ s.p. configuration in the $13/2^+_1$ state of $N = 83$ isotones. Thus, any shell-model interaction with $\varepsilon_{i13/2}>3$ MeV is unlikely to provide such physical two-configuration mixing, and a robust upper limit of $\varepsilon_{i13/2}<3$ MeV is estimated. Assuming that the $i_{13/2}$ s.p. state in $^{133}$Sn is unbound, the physical $\varepsilon_{i13/2}$ value is limited to be 2.4-3 MeV, with is consistent with systematics of the $13/2^+_1$ levels in $N = 83$ isotones.

The authors gratefully acknowledge fruitful discussions with Prof. Y. M. Zhao and Dr. J. M. Allmond, and the technical support from Dr. H. Jin. The authors also thank Dr. C. Qi for providing CD-Bonn matrix elements. This work is supported by the National Natural Science Foundation of China under Contracts No. 11305151, No. 11305101, and No. 11247241. One of the authors (J.H.) thanks the Shanghai Key Laboratory of Particle Physics and Cosmology for financial support (Grant No. 11DZ2260700).

\end{document}